\documentclass[conference]{IEEEtran}
\IEEEoverridecommandlockouts
\usepackage{cite}
\usepackage{amsmath,amssymb,amsfonts}
\usepackage{graphicx}
\usepackage{textcomp}
\usepackage{xcolor}
\def\BibTeX{{\rm B\kern-.05em{\sc i\kern-.025em b}\kern-.08em
    T\kern-.1667em\lower.7ex\hbox{E}\kern-.125emX}}
    
\usepackage{dcolumn}
\usepackage{bm}
\usepackage{hyperref}
\usepackage{algpseudocode}
\usepackage{algorithm}
\usepackage{amsthm}
\usepackage{amssymb}
\usepackage{cleveref}

\usepackage{color}
\newcommand{\estimated}{\ensuremath{\bm{\hat{s}}}}
\newcommand{\measured}{\ensuremath{\bm{\tilde{s}}}}
\newcommand{\boundary}[1]{\bm{\tilde{B}}_{#1}}

\def\be{\begin{equation}}
\def\ee{\end{equation}}

\def\bc{\begin{center}}
\def\ec{\end{center}}
\def\bea{\begin{eqnarray}}
\def\eea{\end{eqnarray}}

\newcommand{\subheading}[1]{\noindent{\textbf{#1}}}

\begin{document}

\title{Higher-order signal processing with the\\ Dirac operator 
    \thanks{MTS acknowledges funding from by the Ministry of Culture and Science of the German State North Rhine-Westphalia (``NRW Rückkehrprogramm''). GB acknowledges funding from the Alan Turing-Roche North Star strategic partnership.}
}

\author{\IEEEauthorblockN{Lucille Calmon}
\IEEEauthorblockA{\textit{School of Mathematical Sciences} \\
\textit{Queen Mary University of London}\\
London, United Kingdom \\
m.l.calmon@qmul.ac.uk}
\and
\IEEEauthorblockN{Michael T. Schaub}
\IEEEauthorblockA{\textit{Department of Computer Science} \\
\textit{RWTH Aachen University}\\
Aachen, Germany \\
schaub@cs.rwth-aachen.de}
\and
\IEEEauthorblockN{Ginestra Bianconi}
\IEEEauthorblockA{\textit{School of Mathematical Sciences} \\
\textit{Queen Mary University of London}\\
London, United Kingdom \\
ginestra.bianconi@gmail.com}
}

\maketitle

\begin{abstract}
The processing of signals on simplicial and cellular complexes defined by nodes, edges, and higher-order cells has recently emerged as a principled extension of graph signal processing for signals supported on more general topological spaces. However, most works so far have considered signal processing problems for signals associated to only a single type of cell such as the processing of node signals, or edge signals, by considering an appropriately defined shift operator, like the graph Laplacian or the Hodge Laplacian. Here we introduce the Dirac operator as a novel kind of shift operator for signal processing on complexes. We discuss how the Dirac operator has close relations but is distinct from the Hodge-Laplacian and examine its spectral properties. Importantly, the Dirac operator couples signals defined on cells of neighboring dimensions in a principled fashion. We demonstrate how this enables us, e.g., to leverage node signals for the processing of edge flows.
\end{abstract}


\section{Introduction}
The processing of non-Euclidean data such as graphs has received a large interest over the last years\cite{bronstein2017geometric,ortega2018graph}.
Recently, there has been a surge of interest in generalizing these ideas to topological spaces modelled by simplicial and cellular complexes~\cite{barbarossa2020topological,millan2020explosive,schaub2020random,grady2010discrete}, thereby enabling the treatment or a richer set of (geometrically supported) signals that are not necessarily supported on the vertices of a graph, but associated to space elements modeled by different kinds of (geometric) cells.

In many cases, we have in fact not only access to signal measurements of a single type, such as edge-flow measurements, but to different types of measurements simultaneously.
For instance, in many physical scenarios we can conduct measurements associated to different types of (geometric) space elements simultaneously, such as measurements associated to points in space as well as to line-segments connecting two such points.
An illustrative example is a resistor network in which we may have access to potential (voltage) measurements associated to certain points, as well as current measurements along certain wires.

Crucially, these types of signals will often not be independent but correlated, or even directly coupled by some kind of physical law --- as is the case for (linear) resistor networks in which Ohm's law provides a linear relationship between the currents and voltages.
Naturally, when processing signals of multiple such types, it is thus desirable to process them jointly, e.g., to enforce such relations, rather than treating each measured signal type independently.

\subheading{Contributions}
We propose the use of the Dirac operator \cite{dirac} as a shift-operator for signals supported on simplicial (and cellular) complexes.
The Dirac operator couples signals on cells in a principled fashion and enables the joint processing of signals supported on cells of different cardinality.
We discuss the spectral properties of the Dirac operator and its relations to the Hodge-Laplacian that has often been used in the literature as shift operator for processing simplicial signals.
We finally illustrate the utility of the Dirac operator by considering basic filtering tasks in which the coupling of signals of different type can be exploited for improved filtering performance.

\subheading{Related literature}
The study of higher-order networks has seen a flurry of interest recently in various domains~\cite{bianconi2021higher,battiston2020networks,bick2021higher,torres2021and}.
Thus far most works on processing higher-order topological signals have however concentrated on processing signals associated to a single type of (higher-order) element, e.g., the processing of flows supported on edges of (higher-order) graphs~\cite{Schaub2018b,barbarossa2020topological,Jia2019,Schaub2021,Schaub2021a,Yang2021,sardellitti2021topological,roddenberry2022signal,bodnar2021weisfeiler}.
In this context, the Hodge-Laplacian has been instrumental and predominantly used as a natural shift operator for the processing of signals on (cellular and) simplicial complexes.
The Dirac operator~\cite{dirac,baccini} has been studied in the context of network dynamics~\cite{topsynchr,calmon2022local,giambagli2022diffusion,millan2020explosive}, but thus far has received little attention in the context of signal processing.

\subheading{Outline}
The remainder of this paper is structured as follows: 
In~\Cref{sec:background}, we briefly review some preliminaries from algebraic topology. 
In~\Cref{sec:dirac_def}, we introduce the Dirac operator and study its spectral properties and relations to the Hodge-Laplacian.
We provide some numerical illustrations of the Diract operator's utility for basic filtering tasks on synthetic and real-data in~\Cref{sec:numerical_experiments}.
Finally, we conclude with some remarks and directions for future work.

\section{Preliminaries}
\label{sec:background}

\subheading{Simplicial complexes and simplicial signals}
Given a set of vertices $\mathcal V$, a $k$-simplex $\sigma_k$ is a subset of $k+1$ vertices.
An abstract \emph{simplicial complex} (SC) $\mathcal X$ is a collection of simplices, such that for any $k$-simplex $\varsigma_k \in \mathcal X$, any subset of $\varsigma_k$ is also an element of $\mathcal X$.
Though our results easily generalize to higher-order SCs, in the following we will focus on SCs consisting only of $0$-simplices (vertices), $1$-simplices (edges), and $2$-simplices (triangular faces).
To each such simplex we associate a scalar signal $s_i\in \mathbb{R}$, which we assemble into a signal vector $\bm s \in \mathbb{R}^{N+E+T}$, where $N$ is the number of vertices, $E$ the number of edges and $T$ the number of triangular faces. We use the convention that vertex signals are indexed first, then edge signals and so on.

\subheading{Boundary maps and Hodge-Laplacians}
We can encode the structure of a simplicial complex via the matrix representations $\bm B_k$ of its \emph{boundary maps}, where $\bm B_k$ records relations between the (arbitrary but fixed oriented) $k$ simplices indexing the columns and $k-1$ simplices indexing the rows. 
For instance, $\bm B_1$ is the well known incidence matrix from graph theory.
We may further equip these boundary operators with weightings, corresponding to diagonal matrices $\bm G_k$.
This leads to weighted incidence matrices of the form $\bm{\tilde{B}_k} = \bm G_{k-1}^{1/2}\bm B_k \bm G_k^{-1/2}$, where the choice $\bm G_k=\bm I$ (for all $k$) yields the standard, unweighted setting again.
Note that by construction $\bm{\tilde{B}_k}\bm{\tilde{B}_{k+1}}=0$, i.e., taking twice the boundary yields zero.
These matrices may alternatively be interpreted as the choice of an inner product induced by $\bm G_k^{-1}$ in the space of $k$-simplex signals. 
For more details on these constructions, see, e.g.,\cite{baccini,grady2010discrete}.

Combining this hierarchy of boundary maps in an appropriate way, gives rise to a sequence of so-called (symmetric) \emph{Hodge-Laplacians}:
\begin{equation}
    \bm L_k = \bm L_k^\text{down} + \bm L_k^\text{up} = \bm{\tilde{B}}_k^\top\bm{\tilde{B}}_k+\bm{\tilde{B}}_{k+1}\bm{\tilde{B}}_{k+1}^\top,
\end{equation}
where by convention we set $\bm B_0=\bm 0$.
For unit weights, $\bm L_0$ is just the graph Laplacian and $\bm L_k$ are the standard combinatorial Hodge-Laplacians, while choosing $\bm G_0$ to encode the weighted node degrees yields the normalized graph Laplacian.

\section{The Dirac operator on simplicial complexes}

\subsection{The (weighted) Dirac operator}
\label{sec:dirac_def}
Unlike the Hodge-Laplacians $\bm L_k$, which may be seen as shift operators between signals defined on $k$-simplices, the Dirac operator~\cite{dirac,topsynchr,baccini} acts jointly on the space of simplicial signals.
On simplicial complexes of dimension $d=2$, i.e., including vertices, edges and triangular faces, the (weighted) Dirac operator can be written as:
\begin{equation}
    \bm D=\begin{bmatrix}
        \bm 0& \boundary{1}& \bm 0\\
        \boundary{1}^\top & \bm 0 &\boundary{2}\\
        \bm 0& \boundary{2}^\top &\bm 0
\end{bmatrix}
\end{equation}
where $\boundary{k}$ are the (weighted) boundary matrices of the SC.
The Dirac operator is an indefinite operator that may be considered the ``square root'' of a block diagonal concatenation of the Hodge-Laplacian operators $\bm L_k$, since we have the relationship 
\begin{equation}
    {{\bm D}}^2=\bm{\mathcal L}=
    \begin{bmatrix}
        \bm L_0 & \bm 0& \bm 0\\
        \bm 0 &{\bm L}_{1}&\bm 0\\
        \bm 0& \bm0& {\bm L}_{2}
    \end{bmatrix}.
\label{eq:squared}
\end{equation}
Similar to how one can obtain a normalized Hodge-Laplacian by choosing appropriate weightings for the boundary maps $\boundary{i}$, it is possible to obtain normalized Dirac operators with a spectrum bounded between $[-1,1]$~\cite{baccini}.
In the following, we will assume such a choice for the weights, i.e., consider the \emph{normalized Dirac operator}.

It is further insightful to decompose the Dirac operator as ${\bm D={\bm D}_1+{\bm D}_2}$ where
\begin{equation}
    \bm D_1=\begin{bmatrix}
        \bm 0& \boundary{1}& \bm 0\\
        \boundary{1}^\top & \bm 0 &\bm 0\\
        \bm 0&\bm 0&\bm 0
\end{bmatrix},
\quad
    \bm D_2=\begin{bmatrix}
        \bm 0& \bm 0& \bm 0\\
        \bm 0 & \bm 0 &\boundary{2}\\
        \bm 0& \boundary{2}^\top &\bm 0
\end{bmatrix}.
\end{equation}
Due to the properties of the boundary matrices, it follows that ${\bm D}_1$ and ${\bm D}_2$ satisfy the relation ${\bm D}_1{\bm D}_2={\bm D}_2{\bm D}_1={\bm 0}$. 
Consequently, we have that $\text{im}({\bm D}_2)\subseteq\text{ker}({\bm D}_1)$ and $\text{im}({\bm D}_1)\subseteq\text{ker}({\bm D}_2)$. 
It follows that similar to the Hodge-decomposition~\cite{grady2010discrete}, the following Dirac decomposition can be established for the space of simplicial signals $\mathcal S \cong \mathbb{R}^{N+E+T}$:
\begin{equation}
    \mathcal{S}=\text{im}({\bm D}_1)\oplus\text{im}({\bm D}_2)\oplus\text{ker}({\bm D})
    \label{Dirac_dec}
\end{equation}
where $\text{ker}({\bm D})$ is given by $\text{ker}({\bm D})=\text{ker}({\bm L}_{0})\oplus \text{ker}({\bm L}_{1})\oplus \text{ker}({\bm L}_{2})$ with a dimension equivalent to the sum of the Betti numbers of the SC.
It follows that a simplicial signal $\bm s$ defined on vertices, links and triangles  can be decomposed uniquely as $\bm s=\bm s_1+\bm s_2+\bm s_{\text{harm}}$,
where $\bm s_n \in \text{im}({\bm D}_n)$ and $\bm s_\text{harm}\in \text{ker}(\bm D)$.
Similar to the Hodge-decomposition, the space $\text{im}(\bm D_1)$ may be interpreted as the (joint) space of gradient flows and node potentials, whereas the space $\text{im}(\bm D_2)$ describes the space of curl flows and associated $2$-simplex potentials.

\subsection{Spectral properties of the Dirac operator}
The Dirac decomposition \cref{Dirac_dec} implies that the eigenvectors of ${\bm D}$ associated to nonzero eigenvalues are either eigenvectors of ${\bm D}_1$ or eigenvectors of ${\bm D}_2$ with non-zero eigenvalue.
The matrix $\bm \Phi$ of the eigenvectors of ${\bm D}$ can thus be written as 
\begin{equation}
    \bm \Phi = 
    \begin{bmatrix}
        \bm \Phi_1& \bm \Phi_2 &\bm \Phi_{\text{harm}} 
    \end{bmatrix}.
    \end{equation}
where ${\bm \Phi}_n$ with $n\in \{1,2\}$ is a matrix containing the eigenvectors of $\bm D$ that span $\text{im}({\bm D}_n)$, and $\bm\Phi_{\text{harm}}$ is a matrix of eigenvectors forming a basis for $\text{ker}({\bm D})$. 

Observe that as $\bm D$ and $\bm{\mathcal{L}}$ share the same null-space by~\Cref{eq:squared}, the vectors $\bm \Phi_\text{harm}$ are simply the direct sum of the eigenvectors  (appropriately padded with zeros) associated to the zero eigenvalues of the Hodge-Laplacians $\bm L_0,\bm L_1$ and $\bm L_2$. 
Further, let us denote the reduced singular value decomposition (i.e., containing only nonzero singular values and vectors) of the boundary operators by $\boundary{1} = \bm U_1 \bm \Sigma_1 \bm V_1^\top$ and $\boundary{2} = \bm U_2 \bm \Sigma_2 \bm V_2^\top$, respectively.
We can then write the matrices of (non-normalized) eigenvectors of ${\bm D_1}$ and ${\bm D}_2$ as:
\begin{align}
    \bm \Phi_1 =
    \begin{bmatrix}
        \bm \Phi_1^= &\bm \Phi_1^\pm
    \end{bmatrix}
    =
    \begin{bmatrix}\bm U_{1} & \bm {\bm U}_{1}  \\
        {\bm V}_1  & -{\bm V}_1 \\
        \bm 0&\bm 0
    \end{bmatrix},
    \label{eq:evecs}
    \\
    \bm \Phi_2 =  
    \begin{bmatrix}
        \bm \Phi_2^= &\bm \Phi_2^\pm
    \end{bmatrix}
    =
        \begin{bmatrix}\bm 0&\bm 0\\{\bm U}_2 & \bm {\bm U}_2  \\
            {\bm V}_2  & -{\bm V}_2 \\
        \end{bmatrix},
        \label{eq:evecs2}
\end{align}
which can be proven by direct computation.
Since $\bm U_i = \boundary{i} \bm V_i \bm \Sigma_i^{-1}$, this implies that the eigenvectors $\bm \Phi_i$ can be split into two sets each: (i) eigenvectors $\bm \Phi^=_i$ for which the signals on the $(i\!+\!1)$-simplices and the $i$-simplices are aligned with the action of the boundary operator (corresponding to the first column-blocks for $\bm \Phi_i$, respectively, in~\Cref{eq:evecs,eq:evecs2}) (ii) eigenvectors $\bm \Phi^\pm_i$ for which the signals on the $(i\!+\!1)$-simplices and the $i$-simplices are anti-aligned with the action of the boundary operator (the second block of columns).
In combination with~\cref{eq:squared}, the above discussion implies that the nonzero eigenvalues of the Dirac operator can be computed as $\lambda^{(i)}_j = \pm \sigma_j^{(i)}$, where $\sigma_j^{(i)}$ is the $j$-th the singular values of the $i$-th boundary map $\boundary{i}$.
Specifically, the positive eigenvalues are associated with the eigenvectors $\bm \Phi^=_i$ of aligned simplicial signals, whereas the negative eigenvalues are associated to the anti-aligned eigenvectors $\bm \Phi^\pm_i$.
By picking out the eigenvectors associated to positive (negative) eigenvalues of the Dirac operator, one can thus find appropriate basis vectors for aligned (anti-aligned) simplicial signals.
This aspect can be employed for filtering simplicial signals as we outline in the next section.

\section{Numerical Experiments}
\label{sec:numerical_experiments}

\subheading{Setup}
We consider a basic linear filtering task, in which we observe a signal of the form
\begin{equation}
    \measured = \bm s + \bm \epsilon,
\end{equation}
Here $\bm s$ is a unit norm signal supported on the simplicial complex we wish to reconstruct and $\bm \epsilon$ is a corresponding noise vector.
To filter these signals we consider infinite impulse response filters (IIR) $\bm H$ filters of the form $\bm H_\gamma = (\bm I + \gamma\bm Q)^{-1}$.
To illustrate how the Dirac operator enables a joint processing of simplicial signals defined on different $k$-simplices, we use suitably chosen positive definite matrix polynomials of the Dirac operator(s) for $\bm Q$, which enables us to tune the filtering behavior of the aligned and anti-aligned eigenmodes in unison.
\begin{equation}
    {\bm Q}= \sum_{j} a_j{\bm D}^j_1 + b_j \bm D_2^j,
\end{equation}
Following from the Dirac decomposition we may consider the action of $\bm D_1$ and $\bm D_2$ independently and thus consider for simplicity only the two filter variants $\bm Q_1(z) = \bm D_1^2 -z\bm D_1^3$ and $\bm Q_2(z)= \bm D_2^2 -z\bm D_2^3$.
Note that due the bounded spectrum of the normalized Dirac operator, for $|z|< 1$ the resulting matrix $\bm Q$ is positive semi-definite.
Note that for positive values of $z$, this form of filter suppresses more strongly the anti-aligned eigenvectors $\bm \Phi_i^\pm$ of the Dirac operator associated to negative eigenvalues and thus promotes aligned signals.
We call such a filter an \emph{aligned filter} in the following.
Similarly, for negative values of $z$ the eigenvectors $\bm \Phi_i^=$ of the Dirac operator associated with positive eigenvalues are strongly suppressed by the filter.
We hence call the filter \emph{anti-aligned}.
From~\Cref{eq:squared}, we can further see that for $z=0$ the filter becomes block-diagonal and thus signals on different types of simplices are effectively treated independently according the respective Hodge-Laplacian $\bm L_k$, which implies that neither an anti-aligned, nor an aligned signal symmetry is promoted.
We call such a filter \emph{uncoupled}.

Note that these filters can also be interpreted as the optimal linear filters in the context following optimization problem:
\begin{equation}
    \min_{\estimated} \left \lbrace \|\estimated -\measured \|_2^2 + \gamma \estimated^\top \bm Q \estimated\right \rbrace \implies \estimated = \bm H_\gamma \measured, 
\end{equation}
where the (positive semi-definite) matrix $\bm Q$ acts as regularizer. 

\begin{figure}[tb!]
\centering
\includegraphics[width=0.95\columnwidth]{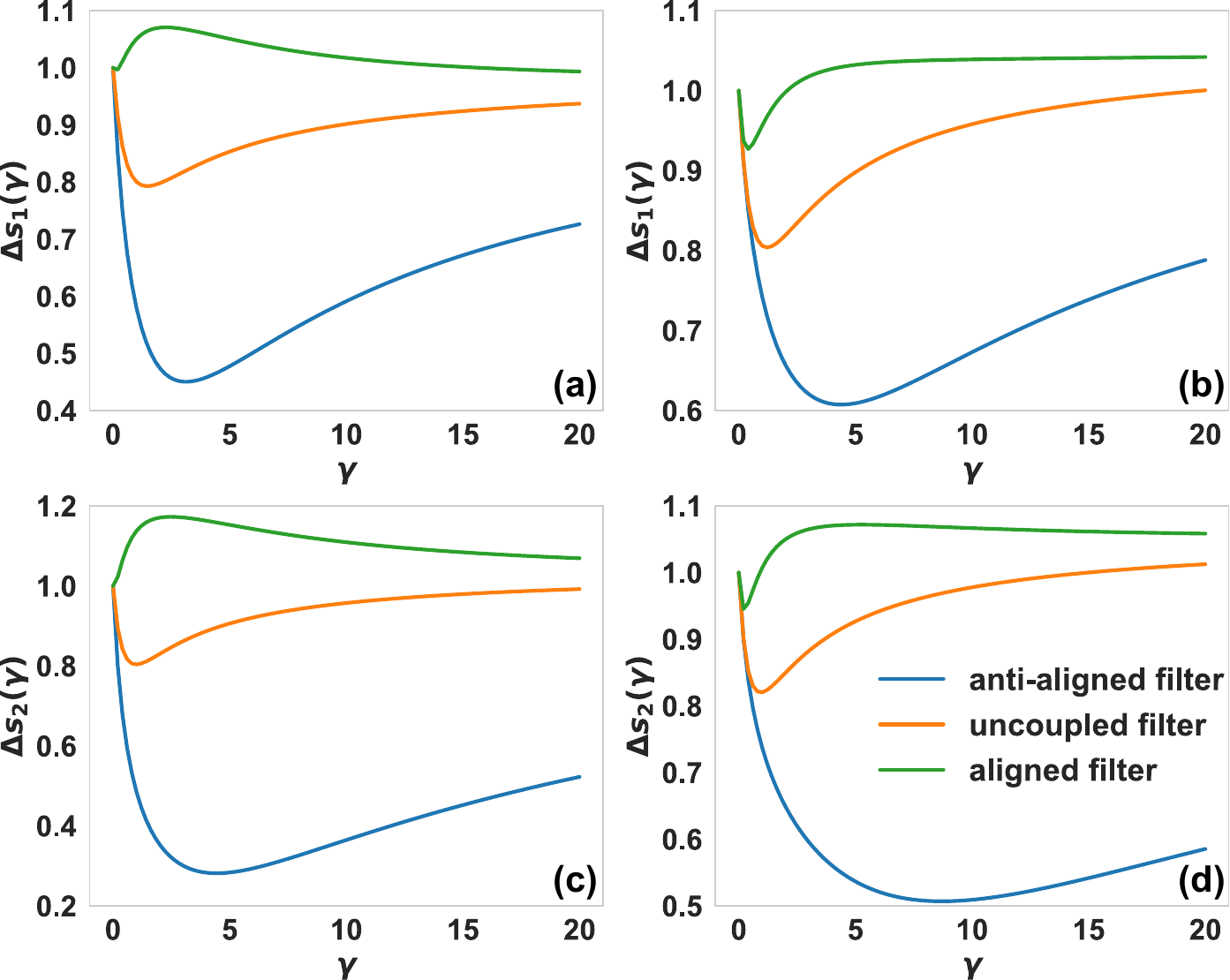}
\caption{\textbf{Dirac filtering promotes compatible signals of simplices of different dimension.}
    We plot the relative error $\Delta s_n(\gamma)$ versus the filter parameter $\gamma$ for an anti-aligned filter ($z=-0.95$), an uncoupled filter ($z=0$), and an aligned filter $z=0.95$) applied to different choices of true signal and noise.  
    All errors shown are averages over $50$ independent realizations.
    \textbf{(a)-(b)} The planted signal is given by eigenvector of $D_1$ associated to the negative eigenvalue largest in magnitude. 
    \textbf{(c)-(d)} The planted signal is given by eigenvector of $D_2$ associated to the negative eigenvalue largest in magnitude.
    In the left column [\textbf{(a), (c)}] the added noise is incompatible with the planted signal symmetry, in the right column [\textbf{(b), (d)}] we use Gaussian noise of (expected) unit norm within the respective subspace.
}
\label{fig:NGF_error}
\end{figure}

\begin{figure*}[tb!]
\centering
\includegraphics[width=1\textwidth]{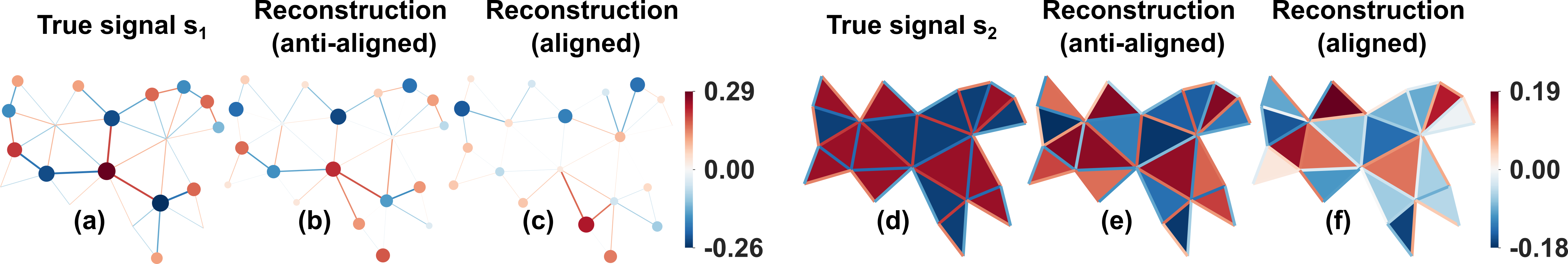}
\caption{\textbf{Illustration of the effects of using an anti-aligned and aligned filter symmetry.} 
    A noisy version of the planted signal $\bm s_1$ (see text), supported on vertices and edges \textbf{(a)} is filtered once  with an anti-aligned filter with parameters $z=-0.95$ \textbf{(b)}, corresponding to the correct signal symmetry, and with an aligned filter with parameters $z=0.95$ \textbf{(c)}, corresponding to a mismatch with the symmetry of the planted signals. Both reconstructed signals are obtained with $\gamma=2.82$. The size of the nodes and the width of the edges encode the value of the signal considered. A noisy version of the planted signal $\bm s_2$ (see text), supported on edges and triangles \textbf{(d)} is filtered once with an anti-aligned filter with parameters $z=-0.95$ \textbf{(e)}, corresponding to the correct signal symmetry, and with an aligned filter with parameters $z=0.95$ \textbf{(f)}, corresponding to a mismatch with the symmetry of the planted signals. Both \textbf{(e)} and \textbf{(f)} are obtained with $\gamma=3.63$.}
\label{fig:NGF_visualisation}
\end{figure*}

\subsection{Application to synthetic data}
We consider an application of our above filtering procedure to signals defined on a simplicial complex, created according to the network geometry with flavor model (NGF) model~\cite{bianconi2017emergent,bianconi2016network}, a comprehensive model of growing simplicial complexes able to generate discrete manifolds with various desirable properties.
Specifically, we consider a simplicial complex formed by vertices, edges and triangular faces, generated with NGF model parameters $s_\text{NGF}=-1$ (flavor) $\beta_\text{NGF}=0$ (inverse temperature).

As the Dirac decomposition implies that signals in $\text{im}(\bm D_1)$ and $\text{im}(\bm D_2)$ are orthogonal and will not interact with each other, we consider two types of signals separately:
\begin{equation}
    \bm s_1 = \phi_1^\pm \in \text{im}(\bm D_1), \qquad \bm s_2 = \phi_2^\pm \in \text{im}(\bm D_2),
    \label{eq:planted_signal}
\end{equation}
where $\bm \phi_1^\pm$, $\bm \phi_2^\pm$ are the eigenvectors associated to the most negative eigenvalues (largest magnitude) of $\bm D_1$ and $\bm D_2$, respectively.
We then add noise to this structured signal $\bm s$ according to two scenarios.

\subheading{Noise with incompatible symmetry}
To showcase the effect of the aligned vs anti-aligned filters, we first consider a scenario in which the noise $\bm \epsilon_n$ added to the signal $\bm s_n$ consists of a random linear combinations of the eigenvectors of the corresponding subspace ${\bm D}_n$ with opposite symmetry relative to the planted signal.
As the planted true signal $\bm s$ (\cref{eq:planted_signal}) has an anti-aligned symmetry, we set the noise to be a random linear combination of the eigenvectors with coefficients drawn from a normal distribution with an expected unit norm in the respective subspace:
\begin{equation}
    \bm \epsilon_n= \bm \Phi_n^=\bm x \quad \text{where} \quad \bm x \sim \mathcal{N}\left(0, \bm I/\sqrt{D_n/2}\right),
    \label{noise_antialigned1}
\end{equation}
where $D_n = |\text{im}(\bm D_n)|$ is the dimension of the subspace spanned by the columns of $\bm D_n$ (i.e., the signal to noise ratio is 1).
Note that this noise thus has the opposite symmetry with respect to the true signal. 

\subheading{Gaussian noise}
A more agnostic choice for the noise $\bm \epsilon_n$ is to assume that it is simply drawn from a multivariate Gaussian distribution within the respective subspace $\text{im}(\bm D_n)$ (normalized to have unit norm in expectation such that the signal to noise ratio is again 1):
\begin{equation}
    \bm \epsilon_n = \bm \Phi_n \bm x \quad \text{where} \quad \bm x \sim \mathcal{N}\left(0,\bm I/\sqrt{D_n}\right)
\end{equation}

\subheading{Results}
In Figure \ref{fig:NGF_error} we assess the errors of the filtered signals $\estimated_n(\gamma)=\bm H_\gamma\bm \measured_n$ in terms the relative errors $\Delta s_n(\gamma)$ 
\begin{equation}
    \Delta s_n(\gamma)=\frac{\|\measured_n-\estimated_n(\gamma)\|_2}{\|\measured_n-\estimated_n(0)\|_2}, \text{ for } n\in \{1,2 \}.
\end{equation}

For both signal $\measured_1$, supported on vertices and edges, and $\measured_2$, supported on edges and triangles, a significant performance gain can be reached for a large range of $\gamma$, when the filter with the right symmetry is chosen, compared to the uncoupled filter. 
This holds both for the illustrative case when the noise is of opposite symmetry as the signal, as well as standard Gaussian noise, which also includes components with the same symmetry as the signal. 
Indeed, our analysis shows that the anti-aligned filter (with $z=-0.95$) performs best and can provide up to $70\%$ error reduction.
In comparison, the uncoupled filter where $z=0$ only provides $20\%$ error reduction at best.
Finally, the aligned filter (with $z=0.95$) results in a reconstructed signal with a larger error than the initially measured (noisy) signal.

An example visualization of the reconstructed signals is shown in Figure \ref{fig:NGF_visualisation}, where it can clearly be seen that the filter with the correct symmetry (here the anti-aligned filter) leads to a better reconstruction, whereas the filter with the wrong symmetry leads to a comparably poor reconstruction of the signal.
Note that due to the frequency response of the filter considered here, the relative errors obtained with the anti-aligned filter increase if the planted signal is spanned by eigenvectors with eigenvalues closer to $0$, especially under Gaussian noise.

\subsection{Application to ocean drifter data}
Beyond synthetic signals built from single Dirac eigenvectors, we test our models on more realistic data.
Specifically, we consider a dataset comprising the trajectories of buoys drifting in the ocean around Madagascar~\cite{schaub2020random}, which may be suitably interpreted as an edge-flow signal defined on a simplicial complex created by triangulating the surface of the earth.
In this case, we do not have direct measurements at the node and triangular faces available. 
However, given the physical nature of the flows, we assume that given the observed flows $\bm \sigma$ the total signal obeys the consistency conditions  $\bm s =\bm \sigma+{\bm D} \bm \sigma$,
which implies that the measured edge-flows are taken as is, and the signals on vertices and triangular faces are reconstructed via $\bm D\bm \sigma$, such that they are aligned (consistent) with the boundary maps appearing in the Dirac operator.
As before we normalize the signal to have unit norm.

\begin{figure}[tb!]
\centering
\includegraphics[width=0.95\columnwidth]{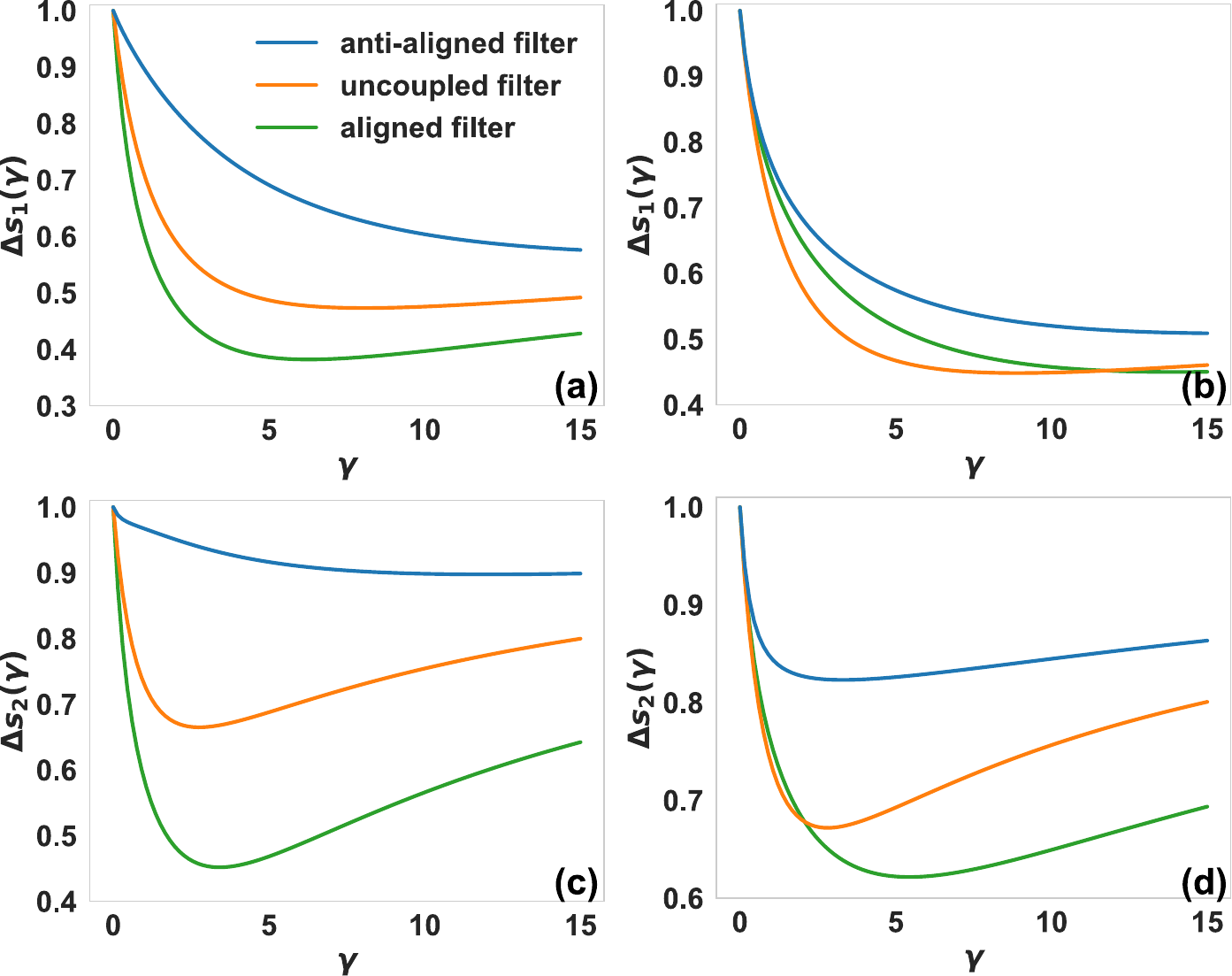}
\caption{\textbf{Performance of Dirac filtering on drifters dataset}
    We plot the relative error $\Delta s_n(\gamma)$ versus the filter parameter $\gamma$ for an anti-aligned filter ($z=-0.95$), an uncoupled filter ($z=0$), and an aligned filter ($z=0.95$) applied to different choices of the planted signal, as described in the text, and the introduced noise.  
    All errors shown are averages over $10$ independent realizations.
    In the left column [\textbf{(a), (c)}] the added noise is incompatible with the planted signal symmetry, in the right column [\textbf{(b), (d)}] we use Gaussian noise of (expected) unit norm within the respective subspace.
    \textbf{(a)-(b)}~The planted signal $\measured_1$ corresponds to the signal component $\measured \in \text{im}(D_1)$.
    \textbf{(c)-(d)}~The planted signal $\measured_2$ corresponds to the signal component $\measured \in \text{im}(D_2)$.
}
\label{fig:buoys}
\end{figure}

We now take the thus constructed signal $\bm \measured$ as our ground truth signal and consider our Dirac filters after we have added either noise with opposite symmetry or Gaussian noise, as described before.
The results are shown in~\Cref{fig:buoys}.
When considering noise with opposite symmetry as in the planted signal (panels \textbf{(a), (c)}), we find up to $60\%$ error reduction can be achieved with the aligned filter ($z=0.95$).
Here the uncoupled filter performs better than could be seen in the case of signal eigenvectors on NGF; for example it achieves approximately $50\%$ error reduction for vertices-edges signals in panel \textbf{(a)}. This can be explained by the fact that the eigenvector decomposition of the true signal is not limited to components with highest eigenvalues, and thus our filter does not match well the frequency profile of the planted signal.
However the aligned filter still shows overall the best performance as it captures the symmetry properties of the signal.

A similar picture emerges when considering the addition of Gaussian noise, as shown in panels \textbf{(b), (d)}, though the filter performance is somewhat worse.
In particular for the signal $\measured_1$ supported on vertices and edges, the decoupled filter performs even slightly better than the aligned filter.
However, for the signal $\measured_2$ supported on edges and triangular faces, the aligned filter still provides a better error reduction compared to the uncoupled and anti-aligned filter.

\section{Conclusions}
We considered the (normalized) Dirac operator as an alternative to the typically considered variants of the Hodge-Laplacian for processing signals on simplicial complexes, and illustrated its utility by means of a few selected example filtering applications.
Specifically, we demonstrated that the Dirac operator enables us to couple signals across simplices of adjacent dimensions and thus facilitates the enforcement of certain compatibility conditions.

However, a number of issues demand future research, e.g., the treatment of more general linear and nonlinear filtering architectures, the extension of these ideas to cellular complexes, or the study of application scenarios.

\bibliography{references}
\bibliographystyle{ieeetr}
\end{document}